\documentclass[letterpaper,12pt]{article}   
\usepackage{spatial} 
\usepackage[draft]{hyperref} 

\begin{document}

\title{The effect of spatial transverse coherence property of a thermal source on Ghost imaging and Ghost imaging via compressive sampling}


\author{Wenlin Gong$^*$, and Shensheng Han}

\address{
Key Laboratory for Quantum Optics and Center for Cold Atom Physics
of CAS, Shanghai Institute of Optics and Fine Mechanics, Chinese
Academy of Sciences, Shanghai 201800, China\\
$^*$Corresponding author: gongwl@siom.ac.cn }

\begin{abstract}
Both ghost imaging (GI) and ghost imaging via compressive sampling (GICS) can nonlocally image an object. We report the influence of spatial transverse coherence property of a thermal source on GI and GICS and show that, using the same acquisition numbers, the signal-to-noise ratio (SNR) of images recovered by GI will be reduced while the quality of reconstructed images will be enhanced for GICS as the spatial transverse coherence lengths located on the object plane are decreased. Differences between GI and GICS, methods to further improve the quality and image extraction efficiency of GICS, and its potential applications are also discussed.\\
\end{abstract}


\maketitle 

In recent ten years, ghost imaging (GI) has attracted lots of attentions in the field of quantum optics \cite{Gatti1}. The image of
an unknown object can be nonlocally reconstructed by the intensity correlation measurements between two light fields.
Both entangled source and thermal light can be used to realize ghost imaging \cite{Gatti1,Gong1,Cheng1,Shen,Pittman,Angelo,Cheng,Gatti2,Bennink,Gong,Zhang1,Zhai,Ferri,Scarcelli}.
The researches of ghost imaging with thermal light have demonstrated that except for incoherence sampling in time domain to be required, the visibility of ghost images also depends on the spatial correlation in space domain of two light fields between the test path and the reference path, distribution of the object, transverse coherence width located on the object plane and so on \cite{Gatti1,Cheng1,Shen,Gong1}. Moreover, the best resolution of recovered images is determined by the sizes of the speckle placed on the object plane based on the previous experimental and theoretical results \cite{Gong,Ferri,Gong2}.

Interestedly, a new imaging approach, called ghost imaging via compressive sampling (GICS), has also proved that an image with high resolution can be quickly and nonlocally obtained by utilizing the sparse prior property of images and using orthogonal measurements \cite{Gong2,Katz,Wang}. Compared with GI, GICS is based on ``global random" measurements and has much higher image extraction efficiency, and the recovered image's resolution is not restricted by the sizes of the speckle placed on the object plane \cite{Gong2}. However, similar to GI, GICS also closely depends on the incoherence property of sampling modality in time domain \cite{Candes,Romberg}, and the effect of sampling modality in space domain on the quality of GICS should be considered. In this letter, the influence of spatial transverse coherence property of light fields on GI and GICS is investigated and their differences are also discussed.

Fig. 1 presents the experimental setup of ghost imaging and ghost imaging via compressive sampling with thermal light. The pseudo-thermal source $S$ is obtained by passing a laser beam, with the wavelength $\lambda$=650nm, into a slowly rotating ground glass disk. The transverse size of the light on the disk $D$ can be controlled by a diaphragm and the light is divided by a beam splitter into a test and a reference paths. In the test path, the light goes through an object then to a detector $D_t$. In the reference path, the light propagates directly to a CCD camera $D_r$. Furthermore, the distances listed in Fig. 1 are as follows: $z$=400mm, and $z_1$=500mm.

By Ref. \cite{Gong,Gong2}, GI can be recovered by measuring the correlation function of intensity fluctuations between two paths
\begin{eqnarray}
\Delta G^{(2,2)} (x_r, y_r) = \int {dx_t}\int {dy_t} \Delta G^{(2,2)} (x_r, y_r; x_t, y_t ) \nonumber\\\sim\int {dx'}\int {dy'} \left| {T(x', y')} \right|^2 \sin c^2 [\frac{D}{{\lambda z}}(x_r  - x')]\sin c^2 [\frac{D}{{\lambda z}}(y_r  - y')].
\end{eqnarray}
where $T(x, y)$ and $D$, respectively, are the object's transmission function and the source's transverse size, $\sin c(x)=\frac{\sin(\pi x)}{\pi x}$ and $\sin c(y)=\frac{\sin(\pi y)}{\pi y}$.

From the results described in Ref. \cite{Gong2}, GICS can be reconstructed by solving the following convex optimization program:
\begin{eqnarray}
\left| {T_{GICS} } \right| = \left| {T'} \right|;{\rm{ \ }} {\rm{ subject \ to: }}{\rm{ \ }} \mathop {\min }\limits_{x,y} {\rm{ \ }}\frac{{\rm{1}}}{{\rm{2}}}\left\| {B_r  - \int {dx} \int {dy} }  {I_r (x,y)\left| {T'(x,y)} \right|^2 } \right\|_2^2  \nonumber\\+ \tau \left\| {\left| {T'(x,y)} \right|^2 } \right\|_1 ,{\rm{ }}\forall _r  = 1 \cdots m.
\end{eqnarray}
where $I_r(x,y)$ is the speckle field placed on the CCD camera $D_r$ plane and $B_r$ denotes the intensity recorded by the test detector $D_t$. $\tau$ is a nonnegative parameter, $\left\| V \right\|_{2}$ denotes the Euclidean norm of $V$, and $\left\| V \right\|_{ 1 }  = \sum\nolimits_i {\left| {\upsilon_i } \right|}$ is the $\ell_1$ norm of $V$. Moreover, $\left| {T_{GICS} } \right|$ is the object's transmission function recovered by GICS reconstruction algorithm.

The numerical simulation and experimental reconstruction results for a double slit transmission plate (with slit width $a$=0.1mm, slit height $h$=1.0mm and center-to-center separation $d$=0.2mm) are summarized in Figs. 2-3, respectively. The different transverse coherence lengths placed on the object plane shown in Fig. 2 and in Fig. 3 can be achieved by changing the transverse sizes of the light on the disk $D$. As $D$ is increased, then the transverse coherence lengths $l_c\approx\frac{\lambda z}{D}$ will be decreased. Fig. 2(a,b) and Fig. 3(b,c), respectively, present the results obtained by GI reconstruction and GICS reconstruction using the same set of measured data. For the GICS reconstruction, we have utilized the gradient projection for sparse reconstruction (GPSR) algorithm \cite{Figueiredo} and $\tau$=0.001 in Eq. (2). From Fig. 2(a) and Fig. 3(b), apparently the signal-to-noise ratio (SNR) of the images recovered by GI reconstruction will reduce whereas its resolution will be improved as the decrease of $l_c$, which is also descried by Eq. (1). Inversely, as $l_c$ is decreased, the quality of the image will be enhanced via GICS reconstruction (Fig. 2(b) and Fig. 3(c)).

To verify the applicability and property of GI and GICS reconstructions for more general images, we have imaged a transmission aperture ($\textbf{SIOM}$). The results for GI and GICS reconstructions in the case of different transverse coherence lengths placed on the object plane, respectively, are presented in Fig. 4(1) and (2) using 2 000 and 1 000 measurements, which also demonstrates the properties and results described in Figs. 2-3.

From the physical point of view, although ghost imaging is a quantum phenomenon and not cased by the statistical correlation of the intensity fluctuations, they are observed in the intensity fluctuations \cite{Scarcelli}. Because the images recovered by both GI and GICS are based on statistical orthogonal measurements to intensity fluctuations in time domain, thus they depend on the properties of incoherence sampling in time domain and the coherence degree of sampling directly influences the sampling efficiency. However, the spatial coherence property of light fields in space domain also has a great effect on GI and GICS. For thermal light, because of the uncertainty relation, pairs of photons having a perfect classical correlation in momentum (or position) cannot exhibit any correlation in position (or momentum) \cite{Bennink}. So as the increase of spatial transverse coherence lengths placed on the object plane $l_c$, the uncertainties of momentum will be decreased and the modes of electromagnetic filed will become much purer, then it is helpful to improve the SNR of GI but the image's resolution will be reduced. Different from GI, the SNR of the images recovered by GICS reconstruction will be enhanced as the spatial transverse coherence lengths $l_c$ is decreased. However, the spatial coherence property of light fields in space domain has no effect on the resolution of GICS, and the resolution of recovered images is closely related to the pixel-resolution of the CCD camera $D_r$ because the intensities located at adjacent pixels of the CCD camera $D_r$ can be resolved due to the nature of compressive sensing reconstruction \cite{Gong2,Herman,Donoho}. Furthermore, from the sparse representation and compressed sensing point of view, the sparsity property of sensing basis (like $I_r(x,y)$) directly determines the image's extraction efficiency and the probability of an image recovered perfectly \cite{Herman}. From Eq. (2), if the spatial transverse coherence lengths $l_c$ is decreased, then the sensing basis coefficients become much more incoherent, so the information of the object ($T(x,y)$) will be embedded in $B_r$ such that the object can be quickly and perfectly recovered with high probability. Thus a classical source, which has high spatial incoherence property of light fields both in time and in space domains, is helpful to enhance the quality of GICS and further improve the image's extraction efficiency, which is very useful to the applications such as the imaging, test and diagnosis of biological tissues with infrared and near infrared light, microscopy, astronomy and so on.

In conclusion, we analyzed the effect of spatial transverse coherence property of thermal light fields on ghost imaging and ghost imaging via compressive sampling. By changing the transverse sizes of the light on the disk to achieve different spatial transverse coherence property of light fields, we demonstrated that the SNR of the images recovered by ghost imaging will be reduced while its resolution will be improved as the decrease of the transverse coherence lengths placed on the object plane $l_c$. However, as the transverse coherence lengths $l_c$ is decreased, the quality of ghost imaging via compressive sampling will be enhanced.

The work was partly supported by the Hi-Tech Research and Development Program of China under Grant Project No. 2006AA12Z115, National Natural Science Foundation of China under Grant Project No. 60877009, and Shanghai Natural Science Foundation under Grant Project No. 09JC1415000.

\textbf{Figure Captions}\\

Fig. 1. Schematic of ghost imaging and ghost imaging via compressive sampling with thermal light.\\

Fig. 2. Numerical simulation results of a double-slit with different transverse coherence lengths placed on the object plane using 300 observations. (a) GI reconstruction; (b) GICS reconstruction.\\

Fig. 3. Experimental reconstruction of the same double-slit (100$\times$100 pixels) using 500 realizations. (a) The correlation of the speckles between two paths when the object in the test path is moved away and the detector $D_t$ is fixed on the object plane; (b) GI reconstruction; (c) GICS reconstruction when the resolution of the CCD camera $D_r$ is 1 pixel. The different transverse coherence lengths placed on the object shown in (1-3) are 276.7$\mu$m, 135.5$\mu$m, and 68.8$\mu$m, respectively.\\

Fig. 4. Experimental results of an aperture ($\textbf{SIOM}$) recovered via GI and GICS reconstructions under the different transverse coherence lengths placed on the object. (a) $l_c$=272.2$\mu$m; (b) $l_c$=193.5$\mu$m; (c) $l_c$=109.6$\mu$m. The images shown in (1-2) are obtained by GI reconstruction (with 2 000 observations) and GICS reconstruction (using 1 000 measurements), respectively.

\newpage

\begin{figure}
\centerline{
\includegraphics[width=8.5cm]{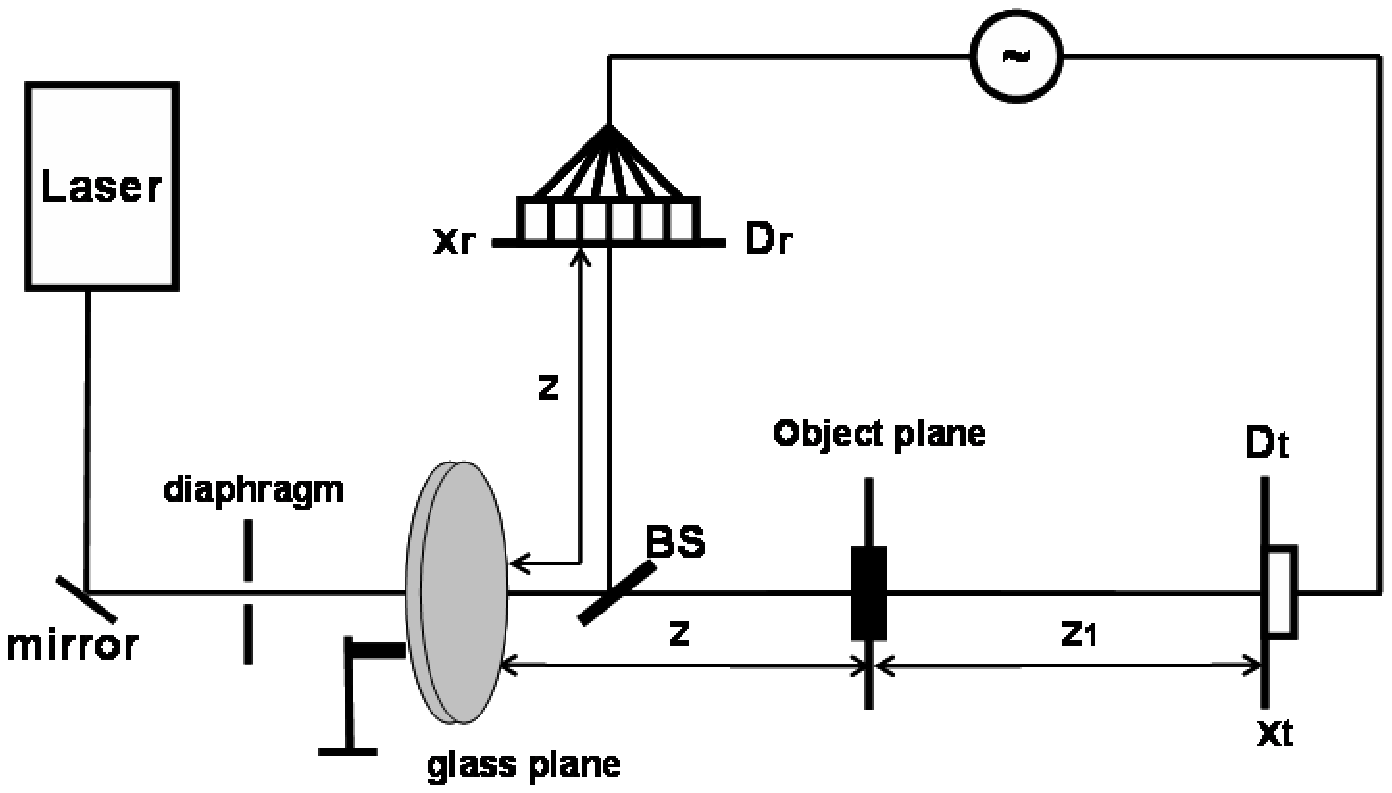}}
\caption{}
\end{figure}

\begin{figure}
\centerline{
\includegraphics[width=8.5cm]{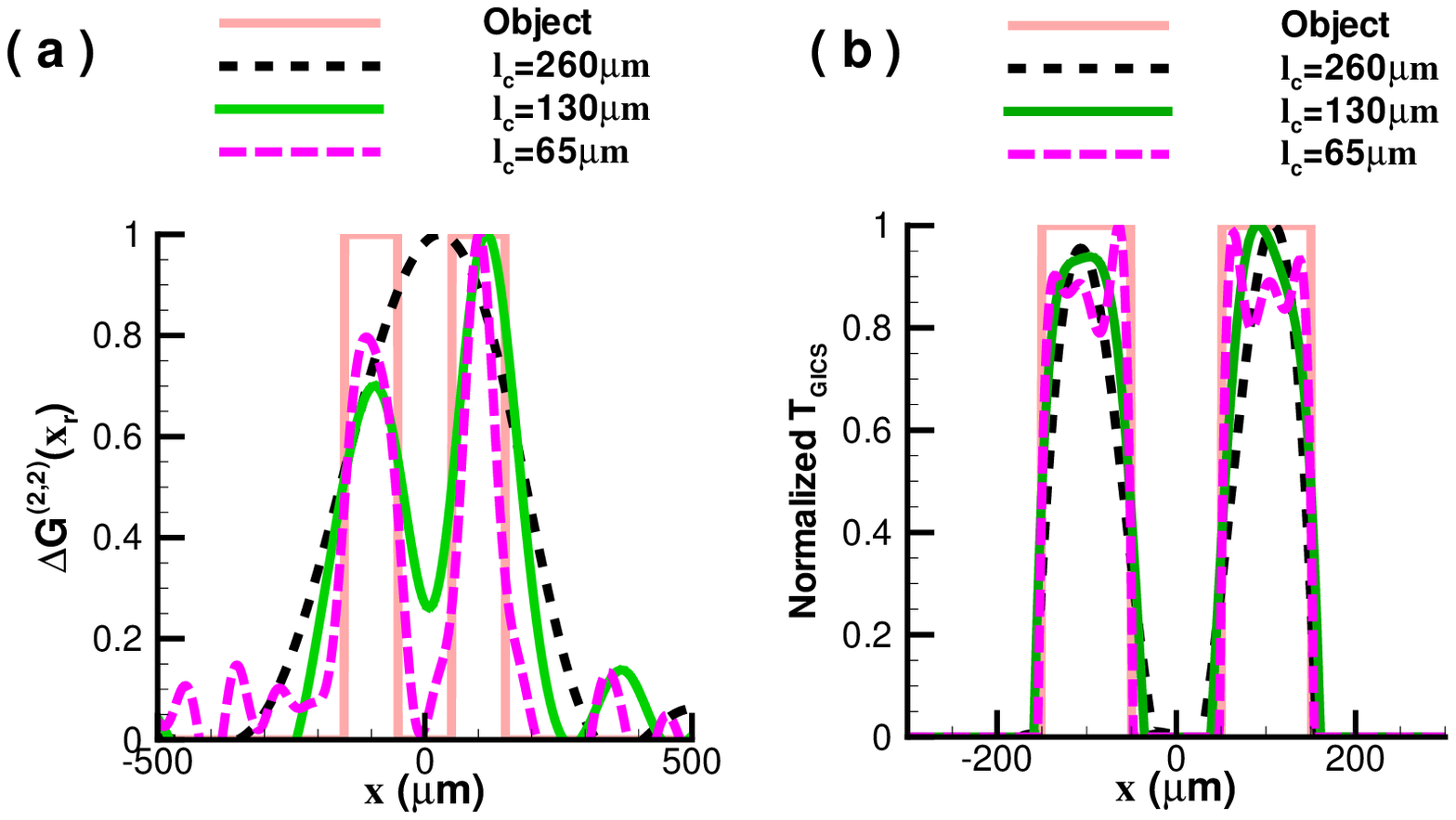}}
\caption{}
\end{figure}

\begin{figure}
\centerline{
\includegraphics[width=8.5cm]{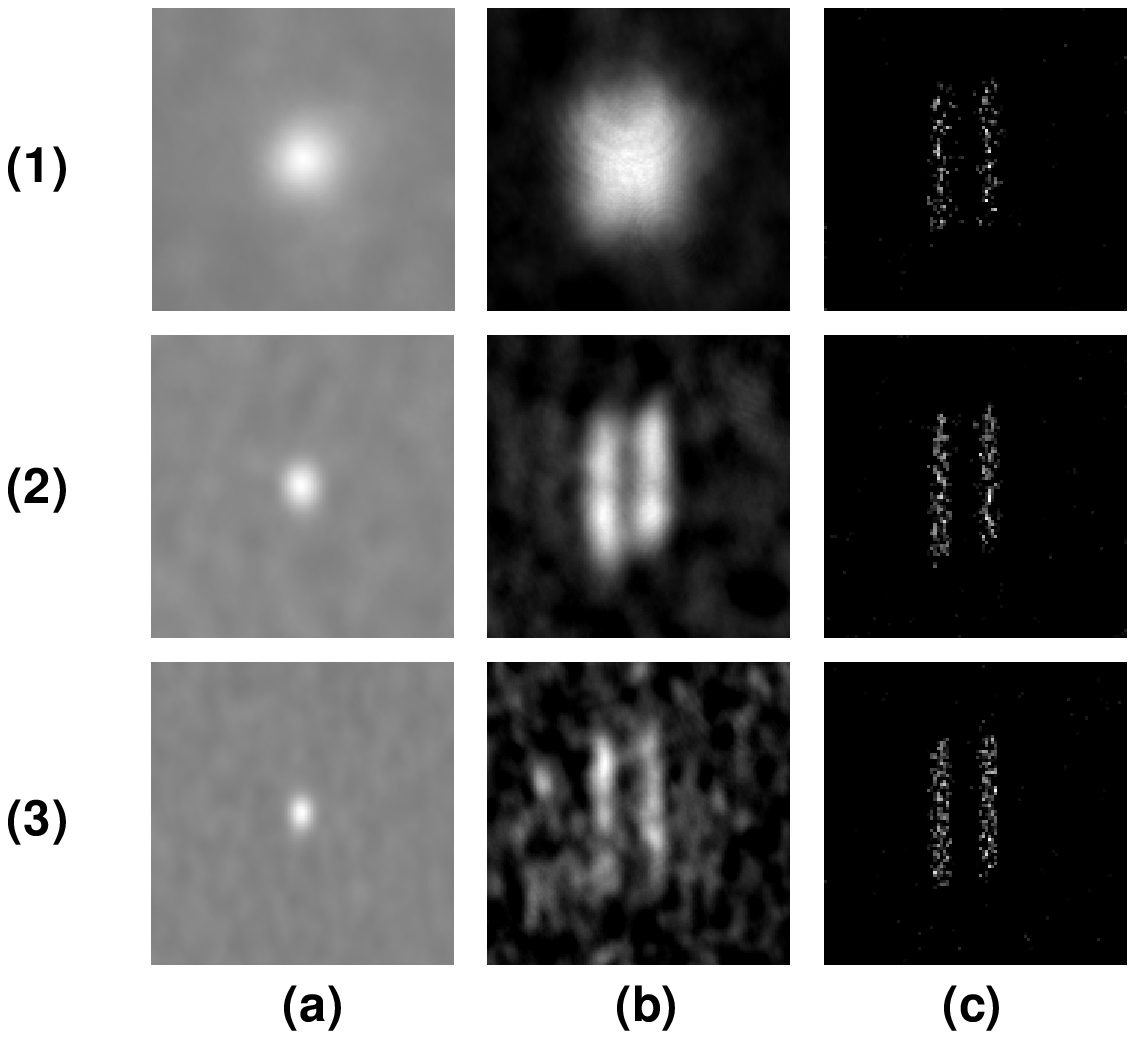}}
\caption{}
\end{figure}

\begin{figure}
\centerline{
\includegraphics[width=8.5cm]{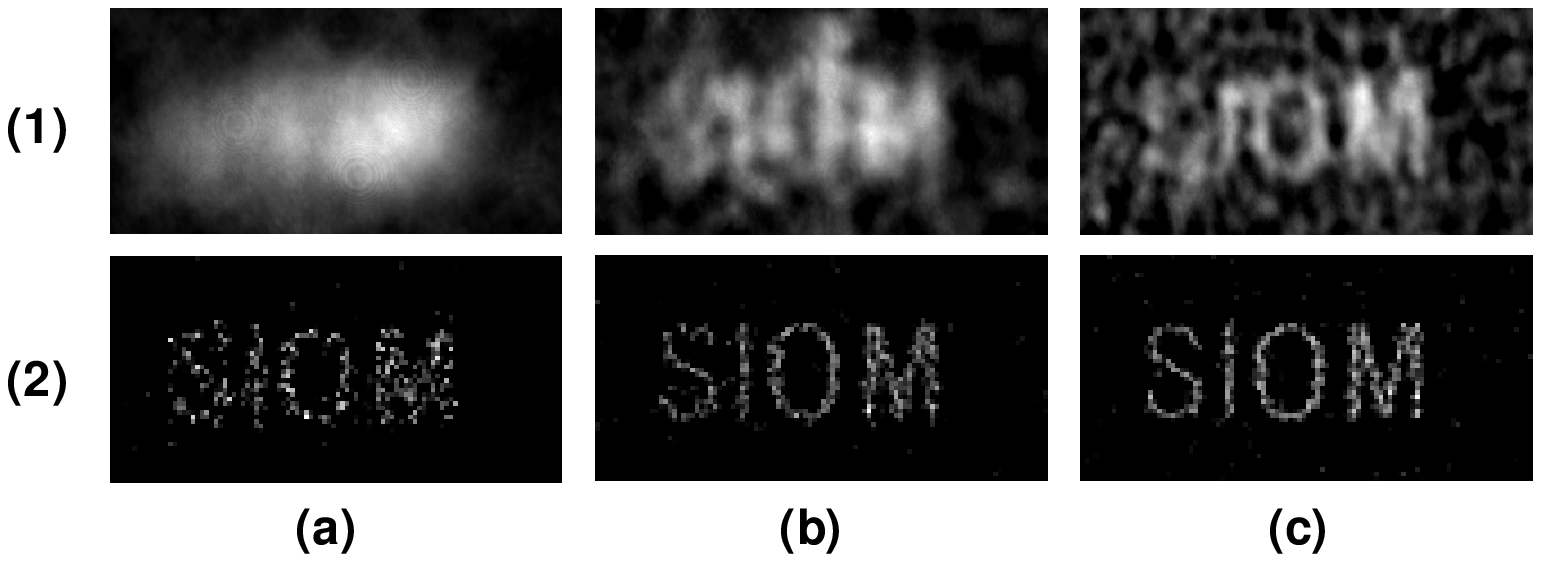}}
\caption{}
\end{figure}

\end{document}